\documentclass[conference]{IEEEtran}
\IEEEoverridecommandlockouts
\usepackage{graphicx}
\usepackage{pgfplots}
\usepackage{circledsteps}
\pgfplotsset{compat=newest}
\usepgfplotslibrary{groupplots}
\usepackage{tikz}
\usepackage{float}
\usepackage{amsmath}

\usepackage{cite}        
\usepackage{amssymb}     
\usepackage{algorithmic} 
\usepackage{array}       
\usepackage{url}         

\usepackage{textcomp}
\usepackage{xcolor}

\def\BibTeX{{\rm B\kern-.05em{\sc i\kern-.025em b}\kern-.08em
    T\kern-.1667em\lower.7ex\hbox{E}\kern-.125emX}}
\begin{document}

\title{Energy-Efficient QoS-Aware Scheduling for S-NUCA Many-Cores
\thanks{This work has received funding from the European Union’s {\em Horizon 2020} research and innovation program for the {\em APROPOS} project under  the  {\em Marie  Skłodowska-Curie} grant  agreement No. 95609.}
}

 \author{
    \IEEEauthorblockN{
        Sudam M. Wasala\IEEEauthorrefmark{1},
        Jurre Wolff\IEEEauthorrefmark{1},
        Yixian Shen\IEEEauthorrefmark{1},
        Anuj Pathania\IEEEauthorrefmark{1},
        Clemens Grelck\IEEEauthorrefmark{1}\IEEEauthorrefmark{2}, and
        Andy D. Pimentel\IEEEauthorrefmark{1}
    }
    \IEEEauthorblockA{
        \IEEEauthorrefmark{1}Informatics Institute, University of Amsterdam, The Netherlands\\
        \IEEEauthorrefmark{2}Institute for Informatics, Friedrich Schiller University, Jena, Germany\\
        Email: s.m.wasala@uva.nl, jurre.wolff@student.uva.nl, \{y.shen, a.pathania, a.d.pimentel\}@uva.nl, clemens.grelck@uni-jena.de
    }
}



\maketitle

\begin{abstract}
Optimizing performance and energy efficiency in many-core processors, especially within Non-Uniform Cache Access (NUCA) architectures, remains a critical challenge. The performance heterogeneity inherent in S-NUCA systems complicates task scheduling due to varying cache access latencies across cores. This paper introduces a novel QoS management policy to maintain application execution within predefined Quality of Service (QoS) targets, measured using the Application Heartbeats framework. QoS metrics like Heartbeats ensure predictable application performance in dynamic computing environments. The proposed policy dynamically controls QoS by orchestrating task migrations within the S-NUCA many-core system and adjusting the clock frequency of cores. After satisfying the QoS objectives, the policy optimizes energy efficiency, reducing overall system energy consumption without compromising performance constraints. Our work leverages the state-of-the-art multi-/many-core simulator {\em HotSniper}. We have extended it with two key components: an integrated heartbeat framework for precise, application-specific performance monitoring, and our QoS management policy that maintains application QoS requirements while minimizing the system's energy consumption. Experimental evaluations demonstrate that our approach effectively maintains desired QoS levels and achieves 18.7\% energy savings compared to state-of-the-art scheduling methods.

\end{abstract}



\section{Introduction}
Ensuring Quality of Service (QoS) in many-core systems is a non-trivial task, exacerbated by resource contention among multiple applications. Traditional metrics used as a proxy for monitoring QoS, such as using Instructions Per Second (IPS), often fail to accurately reflect application-specific performance needs. This inadequacy is particularly evident in contexts like video gaming, where the rendering of each frame constitutes a distinct unit of work. The variability in complexity between frames means that a consistent IPS metric does not necessarily equate to stable frame rates. This disconnect can lead to either inefficient utilization of computing resources or an inability to meet the desired frame rate. Moreover, instruction count-based metrics can be misleading, skewed by idle operations like spin locks. Consequently, there is a growing need for more application-specific performance metrics. Application Heartbeats~\cite{hoffmann2010application}, enabling applications to transparently relay their real-time and target performance, emerge as a compelling alternative, setting the stage for more effective QoS management in many-core systems.

S-NUCA (Static Non-Uniform Cache Access) many-cores~\cite{huh2005nuca}, characterized by their physically distributed yet logically shared last-level cache (LLC), introduce inherent heterogeneity in core access latency. Their LLC access latency varies with the core's proximity to the center of the chip. Cores located closer to the center have, on average, a lower access latency than those cores located at the border.

In S-NUCA systems, there are two main 'knobs' with which QoS can be managed, namely thread migration and DVFS. Thread migration serves to reallocate threads to, e.g., a location closer or further away from the chip's center, whereas DVFS dynamically tailors the cores' voltage and frequency to the performance and power needs. The functionality of these strategies, however, is intimately linked to the individualized demands and behaviors of distinct applications, with generic methodologies potentially leading to suboptimal performance. Thread migration, if performed without an understanding of each application's requirements and the possible contention for resources with other threads, can negatively influence the QoS. Moreover, the diverse ways in which applications respond to DVFS call for a tailored approach to ensure a strategy that is driven by QoS requirements. Previous works~\cite{rapp2020neural,rappnpu} have studied performance optimizations that utilize thread-level performance metrics such as IPS on S-NUCA many-cores; however, they have not studied the effects of an application-level metric such as application heartbeats. 

We propose a management policy that optimizes an application's QoS. It leverages the inherent heterogeneity of an S-NUCA many-core processor. Application-level metrics like Application Heartbeats are important when the application requires execution at a predefined performance rate, such as a stable frame rate. This contrasts with generic applications typically optimized for maximum performance. Therefore, our policy's primary objective is to maintain the Heart Rate (HR) within a predefined target range. We employ a reactive strategy, using DVFS and thread migrations to control the HR. Once the HR stabilizes within the target range, we focus on minimizing the application's energy consumption.

We leverage the open-source and widely-used thermal and performance simulator HotSniper~\cite{pathania2018hotsniper} to test and verify our QoS management policy. The heterogeneity of an S-NUCA many-core strongly depends on the processor chip's physical floorplan. Using a simulator like HotSniper allows us to easily experiment  with different floorplans. Moreover, it enables us to extend our policy to support thermal management in the future. 
We extend HotSniper by introducing a new module that provides heartbeats as a performance metric to its scheduling infrastructure. We use this infrastructure to perform experiments with our QoS management policy.

\textbf{Novel contributions of this paper:}  
\begin{itemize}
    \item We extend the HotSniper thermal and performance simulator by integrating a new module that can be used to simulate applications with integrated Application Heartbeats. This allows us to perform simulations of QoS-aware scheduling algorithms using \emph{Heart Rate} (HR) as a metric.
    \item We develop a reactive QoS management policy that can maintain the HR of an application within a predefined target range, and minimize the energy consumption of the system once the target HR is reached. 
    \item We conduct benchmarking against state-of-the-art techniques, demonstrating that our policy surpasses existing methods in terms of QoS maintenance and energy consumption, confirming its capabilities and potential as a new standard for QoS management in S-NUCA many-cores. 
\end{itemize}

\section{Heartbeat Simulation using HotSniper}
\begin{figure*}[h!]
    \centering
    \includegraphics[width=\linewidth]{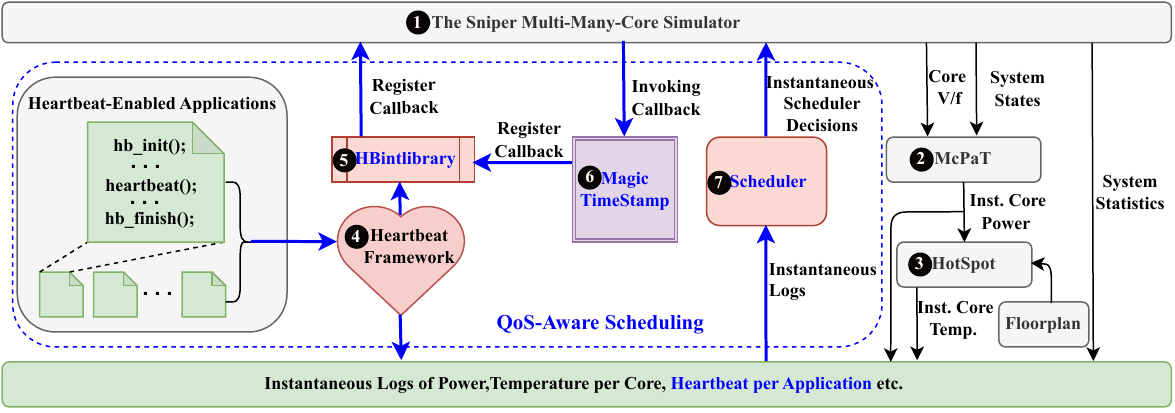}
    \caption{HotSniper tool-flow with heartbeat extensions in blue.}
    \label{fig:heartbeat}
\end{figure*}

\pgfkeys{/csteps/inner color=white}
\pgfkeys{/csteps/fill color=black}

This section details the simulation infrastructure associated with the HotSniper toolchain that is used to simulate S-NUCA architectures.  Moreover, we discuss the integration of the Application Heartbeats framework into this toolchain. Fig. \ref{fig:heartbeat} shows an overview of the HotSniper toolchain and the heartbeat integration. HotSniper employs the Sniper Performance Simulator~\cite{carlson2011sniper} (\Circled{1} in Fig. \ref{fig:heartbeat}) for intricate simulations, monitoring the interactions among critical components such as execution units, caches, and register files. Concurrently, the McPat~\cite{li2009mcpat} (\Circled{2} in Fig. \ref{fig:heartbeat}) tool provides a quantitative analysis of the power consumption by processor cores and memory controllers. The thermal simulator, HotSpot~\cite{huang2006hotspot} (\Circled{3} in Fig. \ref{fig:heartbeat}), uses this data to periodically ascertain the temperatures of each thermal component, all under the guidance of the power model and floorplan configuration. Before discussing the integration of the heartbeats framework into HotSniper, we first provide some background on S-NUCA architectures.

\subsection{Static Non-Unifrom Cache Access (S-NUCA) Architecture}
A Non-Unifrom Cache Access (NUCA) many-core  architecture has a physically distributed but logically shared Last Level Cache (LLC).~\cite{balasubramonian2011multi} Several memory-to-cache address mapping policies have been proposed for many-cores~\cite{das2015exploration}, including Static-NUCA (S-NUCA), which statically interleaves memory addresses across NUCA cache banks at design time. This approach is less flexible at run-time than operating system managed policies like Dynamic-NUCA (D-NUCA). However S-NUCA can be implemented in hardware with minimal overhead.~\cite{kim2011reducing} For example, if we consider a many-core with 64 cores laid out on an 8x8 grid, the LLC would also consist of 64 banks, each bank co-located with a core. 
To access data in the LLC, a thread must reach the appropriate LLC bank associated with the requested memory address via the Network-on-a-Chip (NoC). 
The Sniper simulator (\Circled{1} in Fig. \ref{fig:heartbeat}) can be configured to simulate the performance of a many-core with a S-NUCA architecture.

The latency of LLC accesses depends on the number of hops required on the NoC, quantified by the Manhattan Distance between the core and the LLC bank. Since the interleaving of memory addresses causes threads to access all LLC banks with roughly equal probability, the average LLC access latency for a core is determined by its Average Manhattan Distance (AMD)~\cite{pathania2018task} to all LLC banks. Cores situated near the center of the chip have lower AMD values, resulting in shorter average cache access latencies, while cores towards the edges or corners exhibit higher AMDs and, therefore, longer latencies.


\subsection{Heartbeat Integration in HotSniper}
Optimal Quality of Service (QoS) in complex applications is crucially tied to real-time, application-specific performance measurement, with the heartbeat framework emerging as a vital tool. A heartbeat-enabled application produces a heartbeat every time an iteration of the main loop of the application is executed. The Heartbeat framework (\Circled{4} in Fig. \ref{fig:heartbeat}), encapsulated as a dynamically linked C library, captures these heartbeats and logs the heart rate of the application. HotSniper's scheduler (\Circled{7} in Fig. \ref{fig:heartbeat}) uses these logs, alongside with other performance data to make scheduling decisions.

The successful integration of the Heartbeat framework within the HotSniper toolchain requires overcoming the challenge of accurately representing time due to the discrepancies between real time and the simulator's time. The Heartbeat library, which is crucial for capturing precise timestamps with each heartbeat, encounters a conflict within the Sniper simulator's framework. Sniper runs threads on the native hardware in real time. This causes system calls for timestamps to return the host CPU's clock time, which does not match the slower, simulated timeline Sniper maintains. To solve this, we have developed a so-called Magic Timestamp module (\Circled{6} in Fig. \ref{fig:heartbeat}), which utilizes a callback function (\Circled{5} in Fig. \ref{fig:heartbeat}) to fetch the correct simulation time from Sniper's internal (simulation) clock. Integration with Sniper's magic marker system through a specific C header enables processes and applications to acquire timestamps that accurately reflect the simulated environment's timeline, ensuring accuracy and temporal consistency throughout the simulation's duration.


\section{Related Work}
Effective performance management can be a useful tool for multi- and many-core processors where applications demand specific performance requirements.
Thread scheduling policies for energy and performance optimization on S-NUCA processors have been explored in recent research. Pathania and Henkel~\cite{pathania2018task} proposed a scheduler that exploits the unique topology-based performance heterogeneity of S-NUCA, improving performance by 9.93\% compared to generic schedulers.
Rapp et al.~\cite{rapp2020neural} developed a neural network-based task migration scheduler, optimizing both the timing and location for task migration to enhance performance. Eyerman et al.~\cite{eyerman2011fine} introduced a nuanced 3D fine-DVFS algorithm designed to regulate the frequency and voltage of individual cores, thereby minimizing thermal interference among neighboring cores. Noltsis et al.~\cite{noltsis2019closed} implemented a PID controller to continuously monitor chip temperature and dynamically adjust DVFS in response to temperature fluctuations. Furthermore, Iranfar et al.~\cite{iranfar2020dynamic} crafted a dynamic thermal management (DTM) policy using reinforcement learning, taking into account variables such as fan speed, DVFS settings, and task distribution to optimize overall performance.
Shen et al.~\cite{shen2023thermal} proposes a heuristic based thread migrations policy for thermal management in S-NUCA processors.
Rapp~\cite{rappnpu} explores the use of neural network-based inductive learning for temperature control under QoS constraints, providing nearly optimal decisions with minimal runtime overhead. 

These advancements notwithstanding, there remains a gap in performance management research concerning QoS-aware scheduling that focuses on application level performance metrics such as Application Hearbeats. Hoffmann et al.~\cite{hoffmann2010application} proposed an Application Heartbeat framework that enables applications to report real-time performance data, facilitating accurate QoS tracking. The authors of~\cite{kanduri2018approximation,muthukaruppan2013hierarchical, shamsa2019goal} suggested a scheduling policy for heterogeneous CPUs that leverages program heartbeat data to modulate CPU power consumption and computational accuracy, yielding a more effective balance between power, thermal constraints, and performance, especially when compared to existing methods.  This work highlights the continual evolution of performance management strategies, especially in maintaining QoS objectives. 
More recent efforts~\cite{shen2023thermal2,shen2022tcps,niknam20233d} have explored reinforcement learning and power budgeting techniques for scheduling in multi-core systems, aiming to improve thermal efficiency and real-time performance guarantees. 
However, previous studies focus on big.LITTLE architectures with only two levels of heterogeneity, whereas our proposed QoS management policy targets application-level QoS management on S-NUCA platforms with higher levels of heterogeneity.

 \section{QoS Management Algorithm}
In this section, we discuss the functionality and design of our QoS management policy. The policy aims to optimize application performance and energy efficiency by dynamically adjusting system resources in response to the application's needs. Specifically, it targets three main objectives: 
\begin{enumerate}
    \item achieving the desired Quality of Service (QoS) by maintaining the performance within a predefined target heart rate (HR);
    \item minimizing jitter characterized by fluctuations of HR outside the target range and reducing the overall oscillatory behavior of the HR;
    \item reducing energy consumption once the HR stabilizes within the desired range.
\end{enumerate}

The policy uses clock frequency scaling and thread migrations across the S-NUCA many-core as the knobs to control the performance of the application. By migrating threads to cores with lower AMDs—typically those near the center of the many-core processor—we can reduce the average cache access latency experienced by these threads. This strategic placement minimizes overall LLC latency, which can enhance the HR of applications. Thread migration thus serves as a means to exploit the spatial heterogeneity of S-NUCA architectures, balancing the workload and mitigating resource contention that arises due to non-uniform cache access times.

Clock frequency changes, achieved through dynamic voltage and frequency scaling (DVFS), directly influence the processing speed and power consumption of the cores, offering a more immediate impact on application performance metrics such as HR. Empirical observations have shown that frequency adjustments have a more pronounced effect on HR compared to thread migrations. This insight guides the policy to prioritize frequency scaling when rapid HR adjustments are needed, while still utilizing thread migration to address the longer-term effects of cache access latencies in the S-NUCA architecture.

Designing an effective QoS management policy in this context involves navigating several constraints. The HR metric is inherently application-specific; different applications exhibit widely varying heart rates based on their computational characteristics and workload patterns. Moreover, the HR can change significantly with the number of active threads, adding another layer of complexity to resource management. These factors make it challenging to develop a predictive algorithm that can anticipate HR changes accurately. Therefore, our policy adopts a reactive approach, adjusting resources in response to real-time HR measurements rather than relying on predictive models.

Most applications have application Heartbeats integrated in a way that the main loop of the application consists of a single heart beat for each iteration~\cite{hoffmann2010application}. These applications often use multi-threading to execute these loops, where a number of iterations are assigned to each thread. This means that it is important to measure HR values using a window of a size that is greater than the number of threads. For example, if we assume that there are two threads executing iterations of the same loop in parallel at the same frequency, there is a likelihood of both threads emitting a heart beat ping at roughly the same time.  This will result in a long interval where most of the work is executed, followed by a quick burst of two heartbeats from both threads. Therefore, a window size smaller than the number of threads would result in HR measurements that appear to have large oscillations. Furthermore, if the two threads run at different frequencies it would also create oscillations, which is generally undesirable in this context. As such, to simplify the problem, we maintain a unified frequency for all the cores that run threads from the same application.

Energy consumption is a critical consideration, influenced by both the execution time of applications and the power usage of the cores. Once the HR is within the target range, our policy shifts focus toward minimizing energy consumption without compromising QoS. It carefully manages the trade-off between performance and energy efficiency, ensuring that applications run optimally while conserving power whenever possible. 

\begin{figure}[b]
    \centering
    \includegraphics[width=\linewidth]{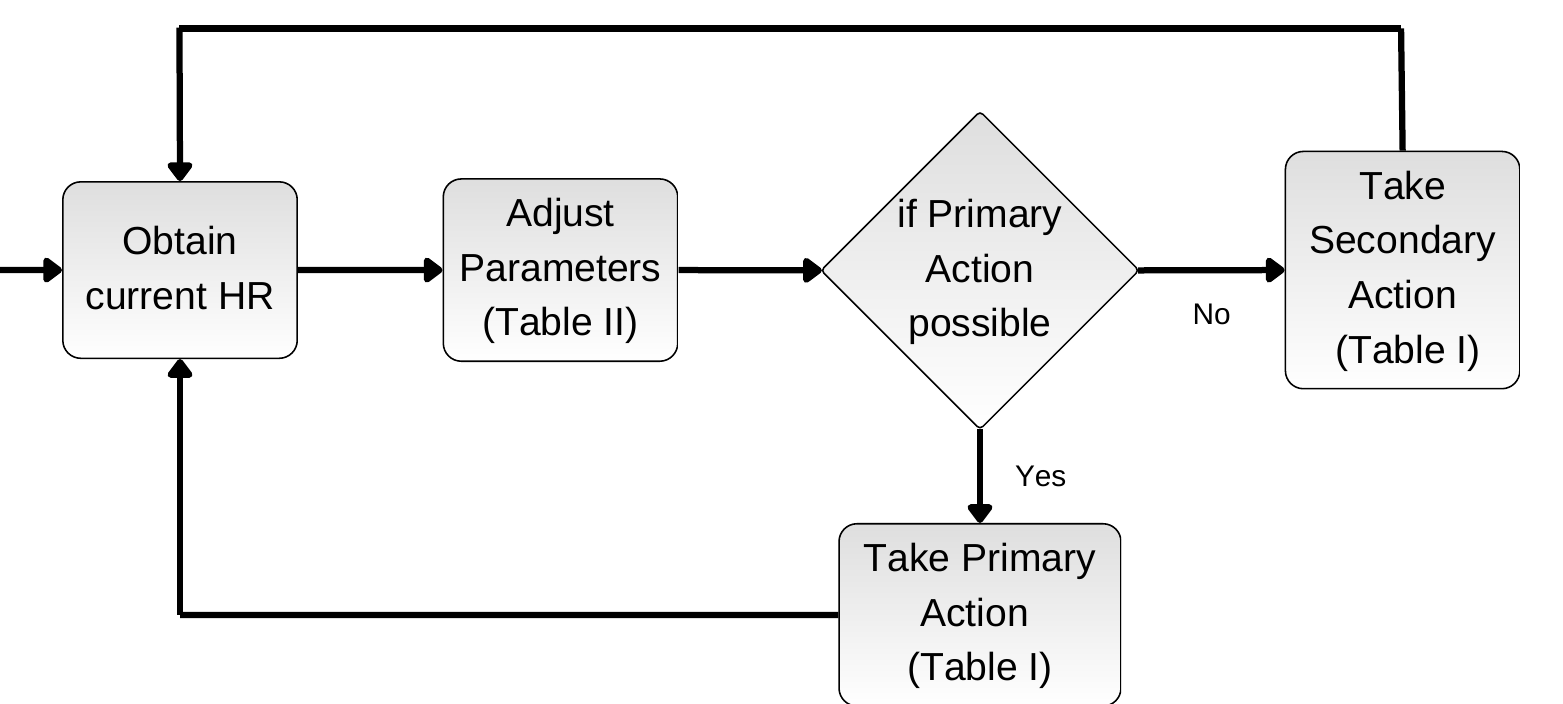}
    \caption{Control flow of the Qos management policy}
    \label{fig:control_flow}
\end{figure}

\begin{figure}[t]
    \centering
    \includegraphics[width=\linewidth]{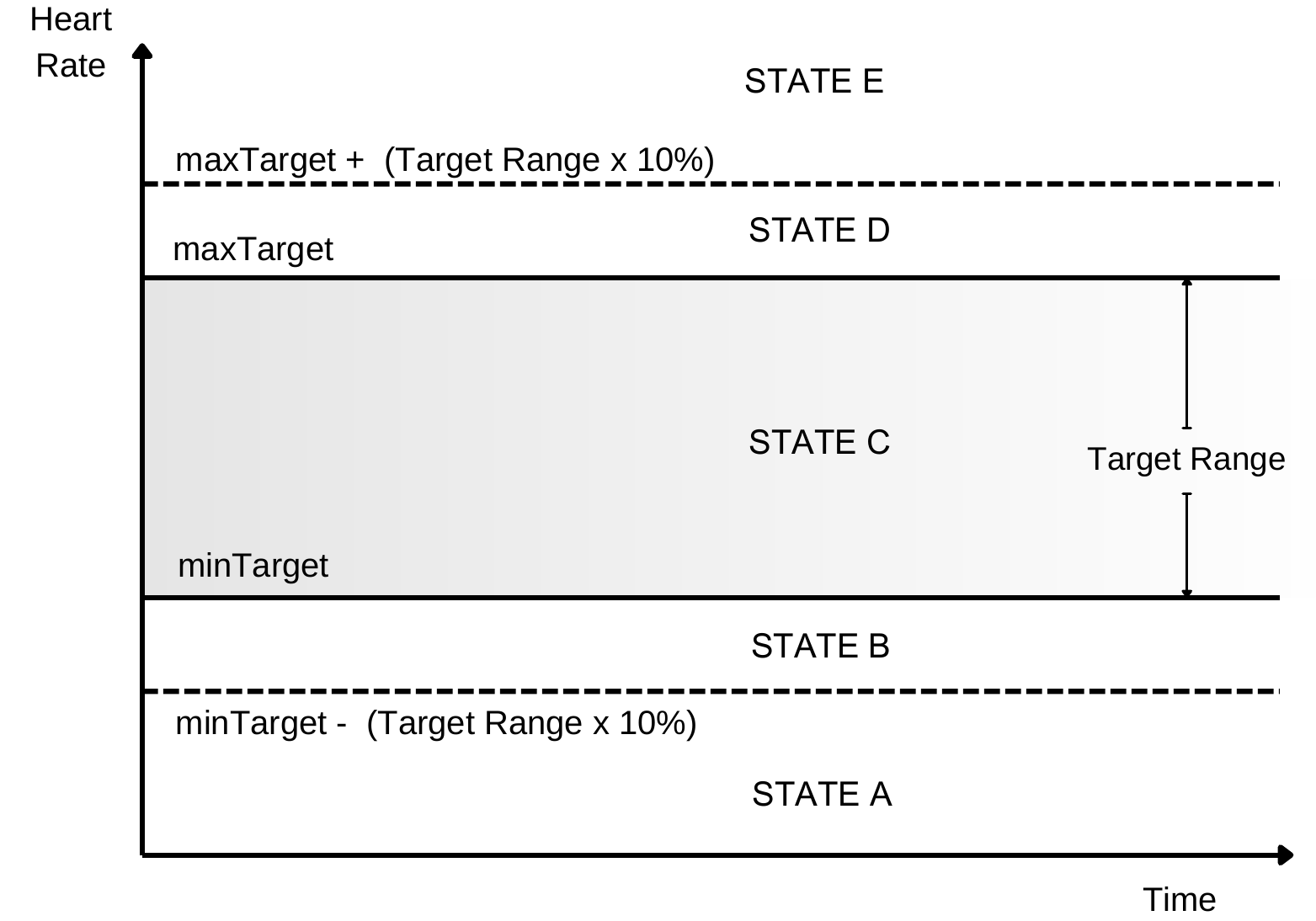}
    \caption{States of the System characterized based on the current HR with respect to the target range.}
    \label{fig:states}
\end{figure}

Fig.~\ref{fig:control_flow} shows an overview of the QoS management policy. At every scheduling epoch, the policy obtains the current Heart Rate (HR) and compares it to the target HR range to determine how close the current HR is to the target. We classify this proximity as a state (Fig. \ref{fig:states}).
Then, the policy adjusts its parameters to prevent overshoot by comparing the current state with the state at the previous epoch. Based on the current state, the policy takes action to bring the HR to the desired range. This action can be thread migration, a large frequency change (macro step), or a small frequency change (micro step). A hierarchical decision-making process determines which of these actions to take, assigning a primary or secondary action based on system conditions. The rest of this section describes this process in more detail.

We characterize the system using five states based on the current Heart Rate (HR) relative to a soft target range. Initially, the soft target range is equivalent to the application's predefined hard target range. However, the policy may later shrink the soft target range to accommodate oscillatory behavior in certain applications. The soft target range always remains within the hard target range. Fig.~\ref{fig:states} shows these five states. States A and E represent a situation where the HR is further away from the target HR range. The policy should attempt to change the heart rate quickly while in these states. States D and B represent a situation wherein the HR is within a pre-defined distance  (10\% in this paper) of the target HR range. Lastly, state C is a situation in which the HR is within the target range, and the policy attempts to minimize energy consumption while ensuring the HR does not go out of the range.

As we have two control knobs, clock frequency and thread migrations, that can be used to alter the performance, we use a hierarchical approach to decide which knob to use. Table \ref{tab:actions} shows which approach the policy uses as the primary and secondary action for each state. The QoS management policy tries to use the secondary action only when the primary action is not possible, i.e. when the frequency has reached the maximum or minimum or when all the threads are at the center or at the corner. For example, if the HR was at state A, the primary action would be to increase the frequency with a macro step, however if the frequency is at the maximum frequency already, the policy will opt for the secondary action, which is to migrate threads towards the center of the many-core.

 Due to frequency adjustments by the policy or due to the oscillatory nature of the application, the HR can sometimes overshoot the target range. When this occurs the policy reactively adjusts several parameters to stabilize the HR, as shown in Table~\ref{tab:adj}. One of the key objectives of the policy is to get the system to state C as quickly as possible. The policy does this primarily by changing the frequency, by a macro step, when the system is in state A or E. However, the main issue with this is that the heart rate can overshoot, potentially causing the system to oscillate between states A and E. To prevent this, we reduce the step size of the frequency change, if such a state change happens. Once inside the target range (state C), the policy needs to ensure that the HR does not wander out of the target range. As discussed above, the HR could show oscillatory behavior with multi-threaded applications. Once inside state C, if the HR value is closer to the target minimum or maximum, this could create situations where the inherent oscillation in the HR could take the HR in and out of the target range. To avoid this, the policy counts the number of times the HR goes out of range and if it exceeds a predefined limit, either the soft minimum target is increased or the soft maximum target is reduced. For the experiments in this paper, we set the predefined limit to 5.

When the HR is within the target, the policy moves to the energy optimization policy. The energy optimization policy is responsible for finding the optimum frequency for energy consumption, while ensuring that the HR does not go out of the soft target range. The issue with energy optimization is that it is difficult to predict at which frequency the energy will be optimal. As the frequency is lowered, less power is consumed, but it takes longer for the same workload to execute, and as such the overall energy consumption is not necessarily reduced. However, as the HR gives an indication of the rate at which the workload is executed, we can use this to predict the effect on overall energy consumption when any frequency change is made. To do this, we increase the frequency by a small amount and observe the change in power consumed compared to the change of HR. If the power and HR ratios, as shown in \eqref{eq:1}, are almost equal, 
then the change in frequency has no effect on the overall energy consumption and the energy consumption is at the minimum. If the HR ratio is greater than the power ratio, more work is done compared to the power increase caused by an increase in frequency. This means that we should keep on increasing the frequency to get better energy performance. If the HR ratio is smaller than the power ratio, the opposite happens and the increase in frequency yields worse energy performance. So we should reduce the frequency to get better energy performance. This reactive approach assumes that the workload is fairly linear so that the power consumption does not vary drastically within a scheduling epoch.

\begin{equation}
    \label{eq:1}
    \frac{HR_{current}}{HR_{previous}} \approx \frac{Power_{current}}{Power_{previous}}
\end{equation}

When changing the frequency, while the HR is inside the target range, we have to be careful not to let the HR go out of range. We use a variable step size where the step size decreases as the HR approaches the minimum/maximum. Equation \eqref{eq:2} shows the variable step size when increasing the frequency $f$.

\begin{equation}
    \label{eq:2}
    Step_{f} = MacroStep_{f}\times \frac{MaxTarget-HR}{MaxTarget-MinTarget} 
\end{equation}

\begin{table}
    \caption{Actions to be taken to control the HR }
    \centering
    \begin{tabular}{p{1cm}|p{3cm}|p{3cm}}
    \hline
    State &               Primary action &             Secondary action \\
    \hline
        A &  Increase frequency by macro step &           Migrate towards center \\
        B &           Migrate towards center &  Increase frequency by micro step \\
        C &     Energy optimization policy &                                - \\
        D &         Migrate away from center & Decrease frequency by micro step \\
        E & Decrease frequency by macro step &         Migrate away from center \\
    \hline
    \end{tabular}

    \label{tab:actions}
\end{table}

\begin{table}[]
    \caption{Parameter adjustment to prevent overshoots}
    \begin{center}
   \begin{tabular}{p{1cm}|p{1cm}|p{5cm}}
    \hline
    Prev State & Curr State &  Adjustment \\
    \hline
    A & D or E & Reduce Macro and Micro stepsize \\
    B & D or E & Reduce Macro and Micro stepsize \\
    C & A or B & mincount +=1; if mincount $ > 5$, increase softMinTarget \\
    C & D or E & maxcount +=1; if maxcount $ > 5$, decrease softMaxTarget \\
    D & A or B & Reduce Macro and Micro stepsize \\
    E & A or B & Reduce Macro and Micro stepsize \\
    \hline
    \end{tabular}
    \label{tab:adj}
    \end{center}
\end{table}

\section{Evaluation}

\begin{figure*}
    \centering
    \includegraphics[]{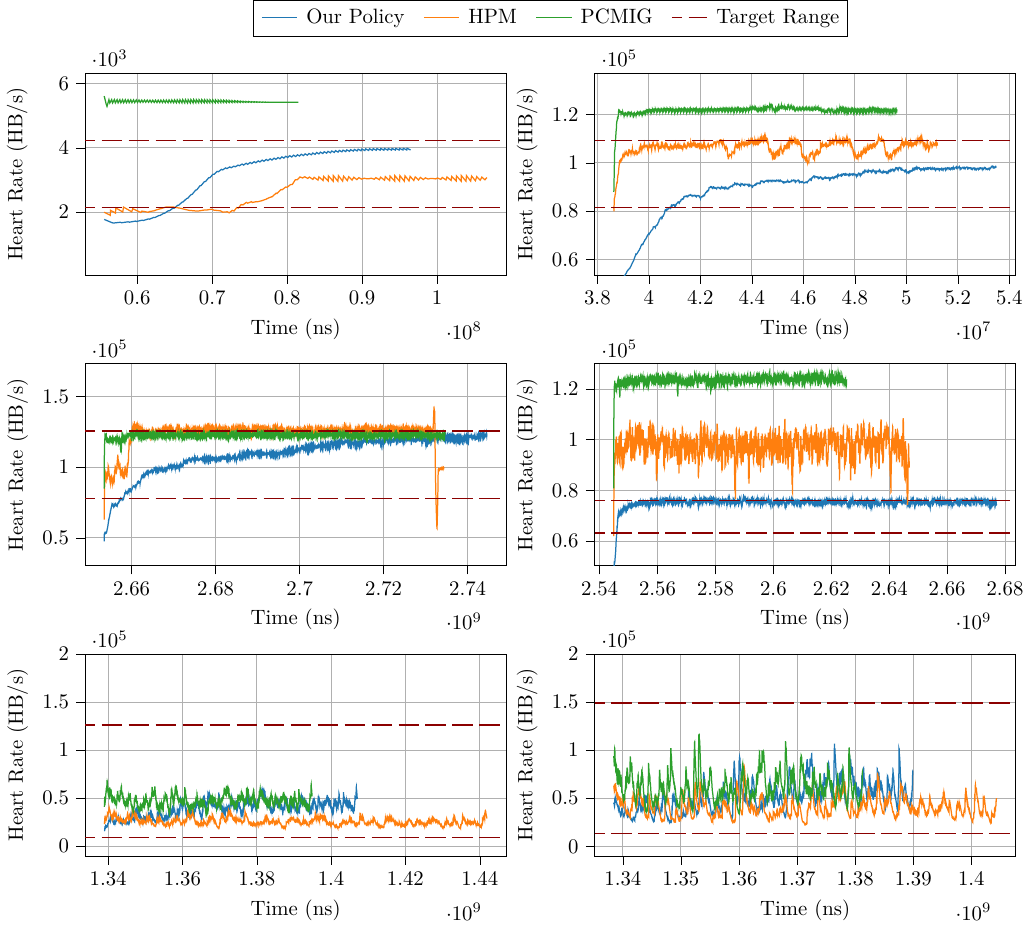}
    \caption{Performance of our policy compared to HPM and PCMiG. (top left: blackscholes (3 threads), top right: blackscholes (15 threads) middle left: canneal (3 threads), middle right: canneal (16 threads), bottom left: dedup (2 threads), bottom right: dedup (10 threads))}
    \label{fig:perf}
\end{figure*}

This section presents the empirical analysis to evaluate the performance of our QoS management policy. 
We use the heartbeat integrated HotSniper simulator to perform experiments on the proposed QoS management policy.
We have selected the \emph{blackscholes},  \emph{canneal}, and \emph{dedup} applications from the PARSEC~\cite{bienia2008parsec} benchmark suite (using simsmall inputs) for our experiments as they are representative for processor-intensive (blackscholes), memory intensive (canneal) and moderate (dedup) workloads.
We assigned a target range for each benchmark and thread count. For each case we simulated their maximum and minimum possible performance conditions to identify the highest and lowest possible HR values. The maximum performance scenario used the highest frequency with threads at the center, while the minimum performance scenario used the lowest frequency with threads at the edges. We then randomly selected two values within this HR range, ensuring they were at least 10\% of the total range apart. These two values constituted the target range for each case.
We simulate a 64-core out-of-order S-NUCA architecture, with an 8x8 grid-based Network-on-Chip (NoC). Both the L1 data (L1-D) and instruction (L1-I) caches have a capacity of 16 KB each. The 8 MB last-level cache (LLC) is divided into 64 banks of 128 KB each. The NoC latency is 1.5 ns per hop, equivalent to 6 CPU cycles at a frequency of 4.0 GHz. The NoC link width is 256 bits.

We use the Hierarchical Power Management (HPM) \cite{muthukaruppan2013hierarchical} and PCMig \cite{rapp2020neural} scheduling policies to compare the performance of our QoS management policy with the state of the art. HPM uses a PID controller to adjust the performance to a predefined target HR. We use an implementation of the HPM policy where the heterogeneity is assigned to a NUCA architecture instead of the big.LITTLE architecture in the original work. PCMig uses a machine learning based approach to predict the performance after a thread migration or frequency change in a NUCA architecture. It optimizes for response time and is not heartbeat aware as it uses IPS as the performance metric.   

\subsection{Evaluation of Performance}
The main objective of the policy is to bring the HR within the predefined target range. Figure~\ref{fig:perf} shows the performance (HR) of the three scheduling policies. For each of the blackscholes, canneal and dedup benchmarks, two executions -- one with a low degree of parallelism (0-5 threads) and one with a high degree of parallelism (10-15 threads) -- are shown. Our approach, reaches the target range in all cases and does not move out of it until the execution ends. As the PCMig algorithm is not aware of the target HR range, it does not produce HRs within the range as expected in most cases. For dedup, PCMig does appear to be in the target HR range, but this is likely by coincidence. The HPM policy gets the HR within the target range sooner than our approach for blackscholes. However, for canneal, HPM appears to overshoot the target range. This may be due to a tuning issue as the PID algorithm's tuning requirements may vary depending on the workload. Our approach shows a more stable HR that remains within the target range compared to HPM. In all cases, our approach shows a significant reduction in oscillations compared to both PCMig and HPM.

\begin{figure}
    \centering
    \includegraphics[]{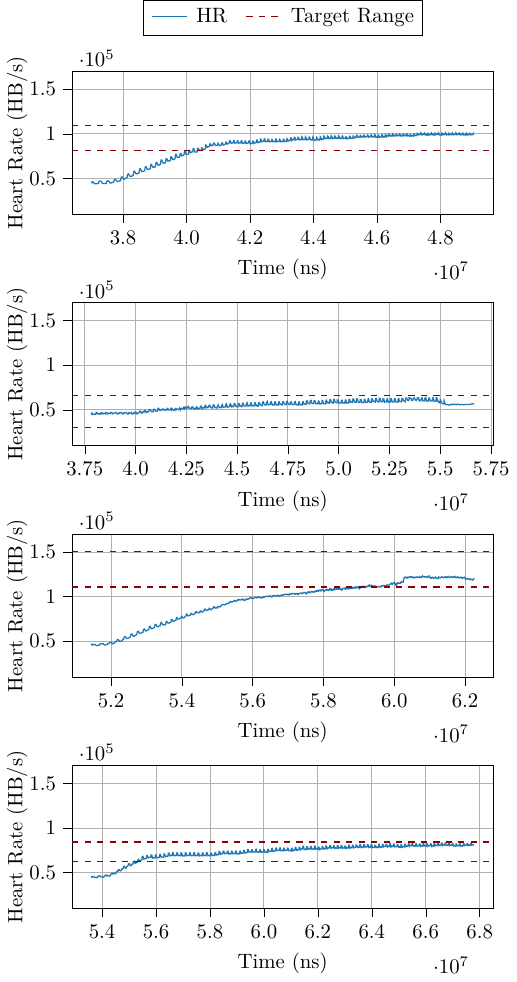}
    \caption{Four parallel blackscholes (12 threads) applications with different target ranges}
    \label{fig:multitask}
\end{figure}

In Figure~\ref{fig:multitask}, we show the performance when concurrently launching four blackscholes applications with 12 threads each. This showcases the policy's capability of handling multiple applications with multiple target HRs.

\subsection{Evaluation of Energy Consumption}
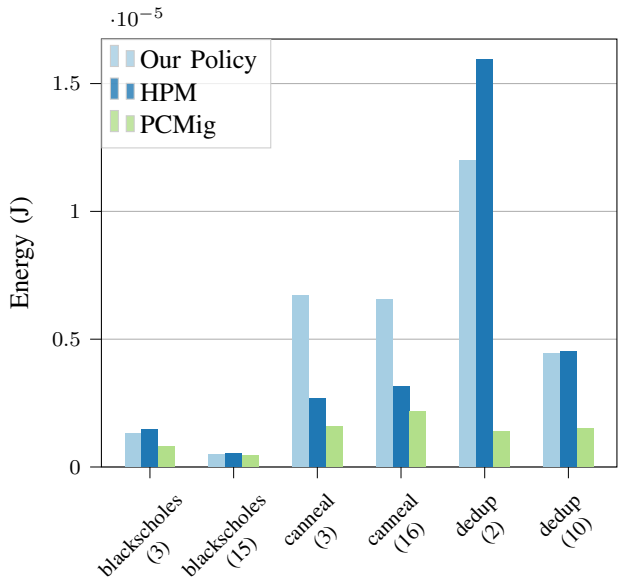
\begin{figure}
    \centering
\begin{tikzpicture}

\definecolor{darkgray176}{RGB}{176,176,176}
\definecolor{lightblue166206227}{RGB}{166,206,227}
\definecolor{lightgray204}{RGB}{204,204,204}
\definecolor{lightgreen178223138}{RGB}{178,223,138}
\definecolor{steelblue31120180}{RGB}{31,120,180}

\begin{axis}[
legend cell align={left},
legend style={
  fill opacity=0.8,
  draw opacity=1,
  text opacity=1,
  at={(0,1)},
  anchor=north west,
  draw=lightgray204
},
tick align=outside,
tick pos=left,
x grid style={darkgray176},
xmin=-0.38, xmax=5.78,
xtick style={color=black},
xtick={0.2,1.2,2.2,3.2,4.2,5.2},
xticklabel style={rotate=45.0},
xticklabels={
  {blackscholes\\(3)},
  {blackscholes\\(15)},
  {canneal\\(3)},
  {canneal\\(16)},
  {dedup\\(2)},
  {dedup\\(10)}
},
ticklabel style={align=center, font=\footnotesize},
y grid style={darkgray176},
ylabel={Energy (J)},
ymajorgrids,
ymin=0, ymax=1.67424102417435e-05,
ytick style={color=black}
]
\draw[draw=none,fill=lightblue166206227] (axis cs:-0.1,0) rectangle (axis cs:0.1,1.29803860222973e-06);
\addlegendimage{ybar,ybar legend,draw=none,fill=lightblue166206227}
\addlegendentry{Our Policy}

\draw[draw=none,fill=lightblue166206227] (axis cs:0.9,0) rectangle (axis cs:1.1,4.75703461414336e-07);
\draw[draw=none,fill=lightblue166206227] (axis cs:1.9,0) rectangle (axis cs:2.1,6.69047576564328e-06);
\draw[draw=none,fill=lightblue166206227] (axis cs:2.9,0) rectangle (axis cs:3.1,6.56304993025882e-06);
\draw[draw=none,fill=lightblue166206227] (axis cs:3.9,0) rectangle (axis cs:4.1,1.19977060770714e-05);
\draw[draw=none,fill=lightblue166206227] (axis cs:4.9,0) rectangle (axis cs:5.1,4.4536793162951e-06);
\draw[draw=none,fill=steelblue31120180] (axis cs:0.1,0) rectangle (axis cs:0.3,1.44818747432648e-06);
\addlegendimage{ybar,ybar legend,draw=none,fill=steelblue31120180}
\addlegendentry{HPM}

\draw[draw=none,fill=steelblue31120180] (axis cs:1.1,0) rectangle (axis cs:1.3,5.2992700408424e-07);
\draw[draw=none,fill=steelblue31120180] (axis cs:2.1,0) rectangle (axis cs:2.3,2.67951689819723e-06);
\draw[draw=none,fill=steelblue31120180] (axis cs:3.1,0) rectangle (axis cs:3.3,3.14812707913022e-06);
\draw[draw=none,fill=steelblue31120180] (axis cs:4.1,0) rectangle (axis cs:4.3,1.59451526111843e-05);
\draw[draw=none,fill=steelblue31120180] (axis cs:5.1,0) rectangle (axis cs:5.3,4.51576088886603e-06);
\draw[draw=none,fill=lightgreen178223138] (axis cs:0.3,0) rectangle (axis cs:0.5,8.10482926122387e-07);
\addlegendimage{ybar,ybar legend,draw=none,fill=lightgreen178223138}
\addlegendentry{PCMig}

\draw[draw=none,fill=lightgreen178223138] (axis cs:1.3,0) rectangle (axis cs:1.5,4.36616012275251e-07);
\draw[draw=none,fill=lightgreen178223138] (axis cs:2.3,0) rectangle (axis cs:2.5,1.57557021944371e-06);
\draw[draw=none,fill=lightgreen178223138] (axis cs:3.3,0) rectangle (axis cs:3.5,2.18909007616172e-06);
\draw[draw=none,fill=lightgreen178223138] (axis cs:4.3,0) rectangle (axis cs:4.5,1.39895782689593e-06);
\draw[draw=none,fill=lightgreen178223138] (axis cs:5.3,0) rectangle (axis cs:5.5,1.49827436911595e-06);
\end{axis}

\end{tikzpicture}
    \caption{Energy consumption of different policies (number of threads are in brackets). PCMig has the best energy performance, however it does not stay within the QoS constraints. HPM for canneal also does not stay within the target range.}
    \label{fig:energy}
\end{figure}

Optimization of the energy consumption is the secondary objective of our QoS management policy, besides the primary objective of maintaining the HR within the required range. Figure~\ref{fig:energy} shows the energy consumption of the three policies. In all cases, PCMig outperforms our approach. PCMig completes the execution of the application in a shorter time, which provides better energy values. However, as discussed above, PCMig is not able to (consistently) keep the application's HR in the required range. In fact, in most cases, the HR of the entire execution is well above the upper limit of the required range.  Our approach produces better energy consumption than HPM in almost all cases, except for canneal where HPM is not within the accepted HR range.

\section{Conclusion}
We presented a reactive QoS management policy for S-NUCA many-core processors, focusing on optimizing performance and energy efficiency. The policy dynamically adjusts system resources to achieve the desired Quality of Service (QoS), measured by application Heart Rate (HR), while minimizing HR fluctuations outside the target range and reducing energy consumption once the HR stabilizes. Using thread migration and dynamic frequency scaling, we exploit spatial heterogeneity by moving threads to optimal cores and adjust processing speeds to control HR effectively. Our reactive approach addresses the inherent challenges posed by the variability of HR across different applications and thread counts. By relying on real-time HR measurements rather than predictive models, our policy adapts swiftly to changing workload characteristics, ensuring that performance targets are met promptly.

Future work includes implementing and testing the QoS management policy on a real system instead of simulations. This will also involve experimenting with benchmarks that had prohibitively long execution times in simulations. Furthermore, our policy currently does not consider thermal constraints. We will explore integrating thermal management strategies into our QoS management policy as future work. 


\bibliographystyle{IEEEtran}
\bibliography{ref}

\end{document}